\documentclass[amsmath, amsfonts, showpacs, twocolumn, prb]{revtex4}
\usepackage{graphicx}
\usepackage{epsfig}
\usepackage{bm}
\usepackage{dcolumn}
\usepackage{amsmath}
\usepackage{amssymb}

\newcommand{\sn}{\mathop{\rm sn}\nolimits}
\newcommand{\cn}{\mathop{\rm cn}\nolimits}
\newcommand{\dn}{\mathop{\rm dn}\nolimits}
\newcommand{\sign}{\mathop{\rm sign}\nolimits}

\begin{document}

\title{Josephson current and density of states in proximity circuits with $s^{+-}$ superconductors}

\author{Stanislav Apostolov}
\affiliation{Department of Physics and Astronomy, Michigan State
University, East Lansing, Michigan 48824, USA}

\author{Alex Levchenko}
\affiliation{Department of Physics and Astronomy, Michigan State
University, East Lansing, Michigan 48824, USA}

\begin{abstract}
We study the emergent proximity effect in mesoscopic circuits that
involve a conventional superconductor and an unconventional pnictide
superconductor separated by a diffusive normal or ferromagnetic
wire. The focus is placed on revealing signatures of the proposed
$s^{+-}$ state of pnictides from the proximity-induced density of
states and Josephson current. We find analytically a universal
result for the density of states that exhibits both the Thouless gap
at low energies, and peculiar features near the superconducting gap
edges at higher energies. The latter may be used to discriminate
between $s^{+-}$ and $s^{++}$ symmetry scenarios in scanning
tunneling spectroscopy experiments. We also calculate Josephson
current-phase relationships for different junction configurations,
which are found to display robust $0$-$\pi$ transitions for a wide
range of parameters.
\end{abstract}

\date{December 1, 2012}


\maketitle

\section{Introduction}

The origin of unconventional superconductivity in ferropnictide
compounds, their phase diagram and symmetry of the underlying order
parameter are topics that attract considerable interest in recent
years, see
Refs.~\onlinecite{Mazin-Review,Paglione-Review,Chubukov-Review} for
reviews. Superconductivity in pnictides emerges in close proximity
to an antiferromagnetically ordered state, and the critical
temperature $T_c$ has a dome-shaped dependence on
doping~\cite{PD-Exp-1,PD-Exp-2} similar to that in cuprates. Due to
their multiband electronic structure with multiple Fermi surfaces
and delicate interplay of interactions in different channels a
number of possibilities for electron ordering are
possible.~\cite{Chubukov-PRB08,Voronstsov-PRB10,Fernandes-PRB10}
Structural transition, competing or coexisting magnetic
spin-density-wave (SDW) and superconducting (SC) orders are being
examples. The latter may be in the form of the conventional
$s^{++}$-wave state that has $s$-wave symmetry in the Brillouin zone
and gaps of the same sign on electron and holes Fermi surfaces.
Alternatively, SC order may appear in the form of an extended
$s^{+-}$ state that looks as $s$-wave from a symmetry point of view
but has opposite signs of the gaps on different sheets of the Fermi
surfaces.~\cite{s-pm-1,s-pm-2,s-pm-3} There may be scenarios of
several SC states with the nodes in the SC gap, of both $s$-wave and
$d$-wave symmetries.~\cite{s-d-1,s-d-2,s-d-3}

Experimentally, the most convincing support in favor of
unconventional symmetry of pnictides is given so far by the observed
spin resonance below $T_c$ in inelastic neutron scattering
measurements on K-doped BaFeAs.~\cite{Christianson-N08} In all the
materials studied, the resonance occurs at the antiferromagnetic
wave vector $Q$ of the parent compound. It is thought to be a
triplet excitation of the singlet Cooper pairs, implying a
superconducting order parameter that satisfies
$\Delta(k+Q)=-\Delta(k)$, which indicates either $s^{+-}$ or
$d$-wave cases. Fabricated $c$-axis Josephson junctions of this
material and ordinary superconductor are suggestive of an $s$-wave
state, but not providing unambiguous evidence for the $s^{+-}$ state
itself.~\cite{Zhang-PRL09} In addition, $ab$-corner-junction
experiments with Co-doped BaFeAs seems to eliminate the option of
$d$-wave pairing.~\cite{Zhou-ArXiv08} Other notable experiments
providing substantial but still indirect support of $s^{+-}$ state
include quasiparticle interference in magnetic field probed by
scanning tunneling microscopy~\cite{Hanaguri-Science10} and
observation of half-integer flux-quantum jumps through the loop
formed by niobium and polycrystalline iron-pnictide
sample.~\cite{Chen-NP10} Finally, there is a growing number of
low-temperatures studies addressing thermodynamics and transport
properties of pnictides, however it is usually hard to deduce
underlying symmetry of a superconductor from such data. For fully
gapped $s$-wave state, one expects to see exponentially suppressed
quasiparticle response and power-law in temperature for the $d$-wave
state with the nodes. The possible ambiguity in interpretation of
data stems from the fact that accidental nodes on the Fermi surface
or impurity-induced subgap states may easily alter low-temperature
behavior of, for example, heat capacity or London penetration depth.

It is widely agreed that a decisive experiment should involve a
phase sensitive probe such as the Josephson effect. Although
original proposals~\cite{Mazin-PRL09,Wu-PRB09} followed mostly
immediately after the $s^{+-}$ candidate symmetry was introduced, no
such direct measurements of the current-phase relationship have been
performed so far for pnictide-based Josephson junctions.
Nevertheless, this inspired a lot of theoretical efforts in finding
simpler geometries or alternative signatures of $s^{+-}$ pairing
state in proximity circuits with pnictides and conventional
superconductors.~\cite{Nagaosa-EPL09,Linder-PRB09,Tsai-PRB09,Chen-PRL09,Ota-PRB10,
Yerin,Berg-PRL11,Koshelev-EPL11,Lin-PRB12,Vakaryuk} A particularly
interesting recent conclusion~\cite{Tsai-PRB09,Koshelev-EPL11} is
that tunneling spectra of weakly coupled $s$-$s^{+-}$ bi-layers
exhibit distinct features characteristic only to sign-changing
symmetry of the gap. Physically, the effect comes from the
frustration in the junction since the gap of an ordinary
superconductor tends to align with one of the gaps of $s^{+-}$
superconductor, and thus becomes in the conflict with the other band
experiencing the anti-proximity effect.

\section{Theoretical framework}

In this work, we consider plethora of effects in
superconductor-metal-superconductor junctions where one or both
superconductors are assumed to have $s^{+-}$ symmetry. The metal is
either normal or ferromagnetic diffusive wire. In the context of the
Josephson effect, in such structures, we find various current-phase
relationships whose shapes depend on the relation between the wire
length and superconducting coherence length, and boundary
transparency. The generic feature is non-analytical behavior of the
current near phase $\pi$, which corresponds to the closing of the
proximity-induced gap in the wire, and robust $0$-$\pi$ oscillations
even without ferromagnets. In the context of the proximity-induced
density of states (DOS) in the wire, we identify fingerprints of the
$s^{+-}$ symmetry, which is thus not only unique to frustrated
$s$-$s^{+-}$ bi-layers.

We build our calculations based on the Usadel
equations~\cite{Usadel} and accompanying Kupriyanov-Lukichev
boundary conditions~\cite{KL} modified for a multiband
case.~\cite{BGK} This quasiclassical theory captures all the
essential features and full complexity of the proximity effect.
Adopting angular parametrization for the normal and anomalous
quasiclassical Green's functions~\cite{Belzig} as
$G=\cos\theta(\omega,x)$ and
$F=\sin\theta(\omega,x)e^{i\chi(\omega,x)}$, Usadel equations take
the form
\begin{subequations}
\begin{equation}\label{Usadel-Eq1}
\partial^2_x\theta-(2\omega/\varepsilon_{Th})\sin\theta=
(\partial_x\chi)^2\sin\theta\cos\theta,
\end{equation}
\begin{equation}\label{Usadel-Eq2}
\partial_x(\sin^2\theta\partial_x\chi)=0,
\end{equation}
\end{subequations}
where $\omega=(2n+1)\pi T$ is Matsubara frequency,
$\varepsilon_{Th}=D/L^2$ is the Thouless energy for the wire of
length $L$, and $D$ is the diffusion coefficient. Spatial
derivatives are taken with respect to the dimensionless coordinate
$x\to x/L$ and we assume quasi-one-dimensional geometry. At the
interface, we have two boundary conditions:
\begin{subequations}
\begin{equation}\label{Boundary-Cond1}
J_\omega=2\sum_{\lambda=1,2}(\delta_\lambda/\gamma_{\lambda})
\sin\theta_B\sin\theta_{s\lambda}\sin\psi,
\end{equation}
\begin{equation}\label{Boundary-Cond2}
\partial_x\theta_B\!=2\!\!\sum_{\lambda=1,2}\!\!
\frac{\cos\theta_{s\lambda}\sin\theta_B}{\gamma_\lambda}\!
\left[\delta_\lambda\tan\theta_{s\lambda}\cot\theta_B\cos\psi
-1\right].
\end{equation}
\end{subequations}
Here, $J_\omega$ denotes the first integral of
Eq.~\eqref{Usadel-Eq2}, index $\lambda$ labels different bands,
$\psi=(\phi-2\chi_B)/2$ and $\phi$ stands for the global
superconducting phase difference across the junction, while factors
$\delta_\lambda=\pm1$ account for the relative shifts of phases
between the bands, and finally parameters $\gamma_\lambda$ represent
dimensionless interface resistances. We also used notations
$\theta_B=\theta(\omega,\pm1/2)$ and similar for $\chi_B$, and
introduced Green's functions of a superconductor in the bulk:
$\sin\theta_{s\lambda}=|\Delta_\lambda|/\sqrt{|\Delta_\lambda|^2+\omega^2}$
and
$\cos\theta_{s\lambda}=\omega/\sqrt{|\Delta_\lambda|^2+\omega^2}$,
with $\Delta_{\lambda}$ being corresponding gaps. Having solved
Usadel equations, one can find a density of states
\begin{equation}\label{N-def}
N(\varepsilon,x)/N_0=\mathrm{Re}[\cos(\omega,x)]_{\omega\to
i\varepsilon}
\end{equation}
upon analytical continuation to real energies, and a Josephson
current-phase relationship
\begin{equation}\label{I-def}
eI(\phi)R_N=2\pi
T\sum_{\omega}J_\omega=\int\tanh\frac{\varepsilon}{2T}\mathrm{Im}J_\varepsilon
d\varepsilon
\end{equation}
upon summation over Matsubara frequencies, where $N_0$
is bare density of states in a metal and $R_N=L/e^2DN_0S$ is normal
state wire resistance of cross-section area $S$.

\section{Density of states}

Consider a symmetric $s|n|s$ junction. In the absence of
superconducting phase difference between the leads we have $\chi=0$
and the whole system of equations simplifies to one:
\begin{equation}
\partial^2_x\theta-(2\omega/\varepsilon_{Th})\sin\theta=0,
\end{equation}
which has to be solved for $x\in[-1/2,1/2]$. Since this is the same
equation as for the nonlinear pendulum it can be integrated exactly
in terms of the Jacobi elliptic functions. Indeed, the above
equation has a simple first integral:
\begin{equation}
(\partial_x\theta)^2=(4\omega/\varepsilon_{Th})[\cos\theta_0-\cos\theta],
\end{equation}
where integration constant $\theta_0=\theta(\omega,0)$ was chosen to
be at the middle of the wire due to obvious symmetry reasons. To
perform a subsequent second integration, we change variables as
\begin{equation}\label{m}
\cos\theta=\frac{2m\cos^2\phi}{1-m\sin^2\phi}-1,\quad
m=\cos^2(\theta_0/2),
\end{equation}
and find
\begin{equation}
x\sqrt{\frac{2\omega}{\varepsilon_{Th}}}=\int^{\phi}_{0}\frac{d\phi}{\sqrt{1-m\sin^2\phi}},
\end{equation}
which is a tabulated integral. Finally, using the Jacobi elliptic
functions $\cn$ and $\dn$, the solution appears in the form
\begin{equation}\label{theta-dos}
\cos[\theta(\omega,x)/2]=\cos(\theta_0/2)\frac{\cn(u,m)}{\dn(u,m)},\quad
u=x\sqrt{\frac{2\omega}{\varepsilon_{Th}}}.
\end{equation}
It is important to keep in mind that the modulus of the Jacobi
functions is actually an energy-dependent function $m(\omega)$, see
Eq.~\eqref{m}. By using now Eq.~\eqref{theta-dos} in the boundary
condition Eq.~\eqref{Boundary-Cond2}, one finds a closed algebraic
equation for the unknown integration coefficient in the form
\begin{eqnarray}
u_B\sqrt{1+m}\frac{\sn(u_B,m)}{\dn(u_B,m)}+
\mathcal{F}\frac{\cn(u_B,m)}{\dn^2(u_B,m)}\nonumber\\=
\frac{\mathcal{G}}{\sqrt{m(1-m)}}
\left[\frac{1}{2}-\frac{1-m}{\dn^2(u_B,m)}\right]\label{Boundary-Cond-u}
\end{eqnarray}
where $u_B=\sqrt{\omega/2\varepsilon_{Th}}$, and
\begin{eqnarray}
\mathcal{G}(\omega)=\sum_{\lambda}\cos\theta_{s\lambda}/\gamma_\lambda=
\sum_\lambda\frac{\omega}{\gamma_\lambda\sqrt{|\Delta_\lambda|^2+\omega^2}},\label{G}\\
\mathcal{F}(\omega)=\sum_{\lambda}\delta_\lambda\sin\theta_{s\lambda}/\gamma_\lambda=
\sum_\lambda\frac{\delta_\lambda|\Delta_\lambda|}{\gamma_\lambda\sqrt{|\Delta_\lambda|^2+\omega^2}}.\label{F}
\end{eqnarray}
Equation~\eqref{Boundary-Cond-u} defines $m$ and thus $\theta_0$ as
a function of energy $\omega$ and together with Eqs.~\eqref{N-def}
and \eqref{theta-dos} it provides a complete analytical solution for
the Green's function in the wire.

\begin{figure}
  \includegraphics[width=8cm]{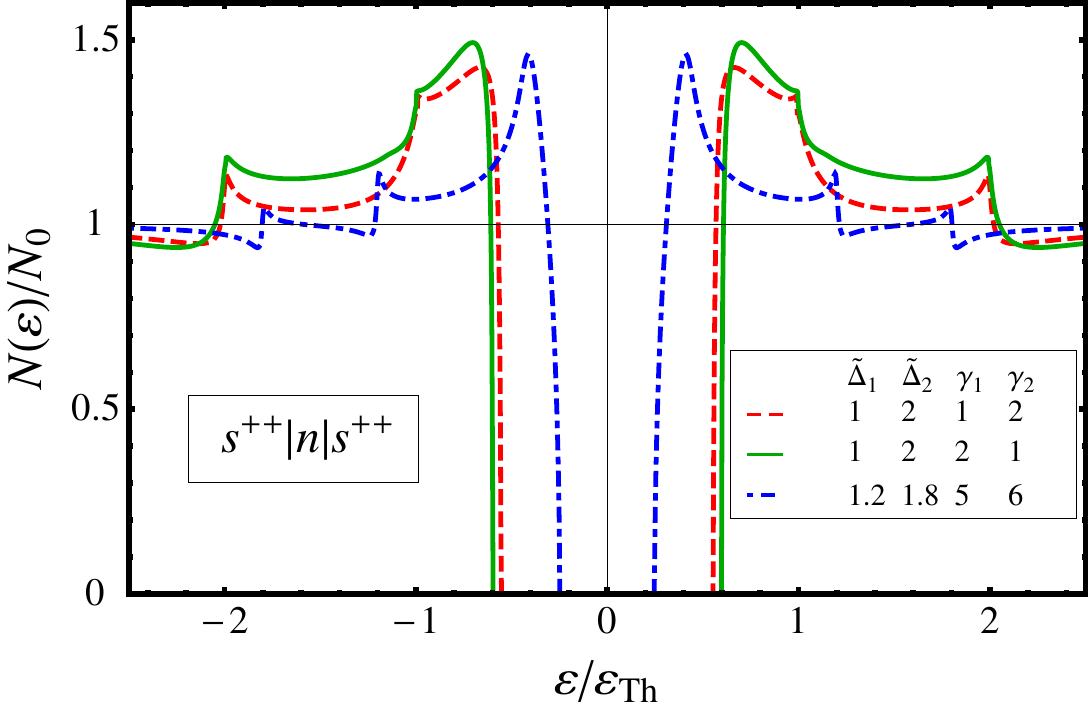}\vspace{-.25cm}\\
  \caption{Representative density of states spectrum in the normal wire as
  induced by the proximity effect between two $s^{++}$ two-band superconductors. Inset shows
  parameters, and bulk gaps were normalized to the Thouless energy
  $\tilde{\Delta}_\lambda=\Delta_\lambda/\varepsilon_{Th}$.}\label{Fig-DOS1}
\end{figure}

\begin{figure}
  \includegraphics[width=8cm]{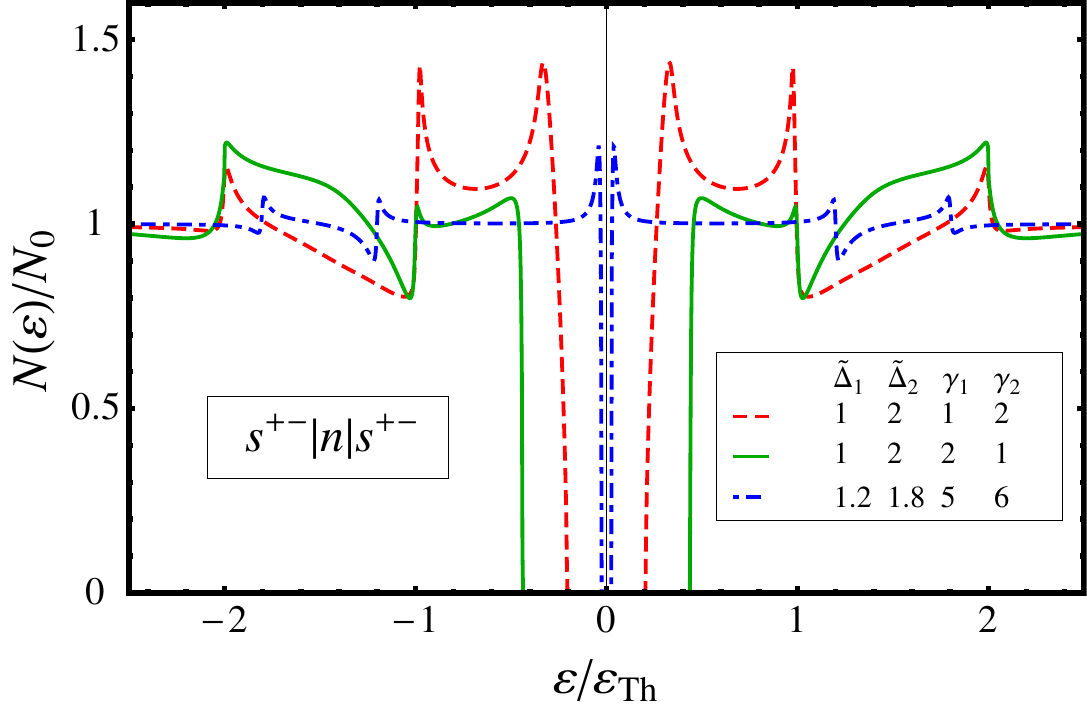}\vspace{-.25cm}\\
  \caption{Representative density of states spectrum in the normal wire as
  induced by the proximity effect between two $s^{+-}$ two-band superconductors.
  The anti-symmetric Fano-like features near the gap edges
  $\tilde{\Delta}_\lambda=\Delta_\lambda/\varepsilon_{Th}$ may provide a definite fingerprint
  of $s^{+-}$-superconductivity.}\label{Fig-DOS2}
\end{figure}

In particular, we can find density of states in the middle of the
wire as $N(\varepsilon)/N_0=\mathrm{Re}[2m(i\varepsilon)-1]$, which
exhibits very rich structure. Indeed, Fig.~\ref{Fig-DOS1} shows
representative profiles of $N(\varepsilon)$ in $s^{++}|n|s^{++}$
junction for different choice of parameters. One finds a
proximity-induced energy gap $\varepsilon_g$ in the spectrum of a
wire, which scales with the Thouless energy
$\varepsilon_g\sim\varepsilon_{Th}$. Asymptotic analysis near the
gap, $\varepsilon-\varepsilon_g\ll\varepsilon_g$, shows that DOS has
a square-root singularity
$N(\varepsilon)\propto\sqrt{\varepsilon/\varepsilon_g-1}$, similar
to that in a single-band $s|n|s$ junctions.~\cite{AL-DOS}
$N(\varepsilon)$ then rapidly grows, passes through the maximum and
has two additional peak-like features at higher energies near the
superconductive band gaps $\Delta_{\lambda}$. This picture has to be
contrasted to the DOS profile in $s^{+-}|n|s^{+-}$ junctions shown
in Fig.~\ref{Fig-DOS2}. The low-energy behavior is similar but the
energy gap is reduced due to anti-proximity effect induced by the
$\pi$-shifted band. The conceptual difference appears near the band
gaps $\Delta_{\lambda}$ where instead of peaks one finds Fano-like
antisymmetric features. This important detail is specific for the
$s^{+-}$ symmetry case and can be looked for in the tunneling
experiments.

\section{Josephson current}

Phase-sensitive measurements are clearly more challenging.
Nevertheless, we develop a theory for the Josephson effect in
mesoscopic $s|n|s^{+-}$ circuits with the idea that some limits
considered here will be useful for the future experiments.

In the presence of a superconducting phase gradient in the wire
finding an analytical solution of Usadel equations represents a
difficult technical problem. In the limit of the long junctions
however $L\gg\sqrt{D/T}$ calculation of the Josephson current
simplifies considerably. In this case, it is possible to neglect the
mutual role of superconducting leads and introduce an ansatz for the
anomalous Green's function
$F=e^{i\phi/2}\sin\theta^R+e^{-i\phi/2}\sin\theta^L$, where the
functions $\theta^{R(L)}$ satisfy the same sin-Gordon equation as in
the case of DOS calculations. Solving it separately near right
(left) boundary for $\theta^{R(L)}$ respectively we
find~\cite{Zaikin}
\begin{equation}
\tan[\theta^{R(L)}(x,\omega)/4]=\mathcal{B}_{R(L)}(\omega)
\exp[\pm(x\mp 1/2)L/\xi_\omega]
\end{equation}
where we introduced coherence length $\xi_\omega=\sqrt{D/2\omega}$.
This approximation conserves the current in the normal layer with
the exception of the narrow region of the order $\xi_{\omega=T}$
near the boundaries. The two integration coefficients
$\mathcal{B}_{R(L)}$ are to be found from the boundary conditions
Eq.~\eqref{Boundary-Cond2} at both interfaces, which can be reduced
to the algebraic equation:
\begin{eqnarray}
4\mathcal{G}_\alpha(\mathcal{B}_\alpha-\mathcal{B}^3_\alpha)-
\mathcal{F}_\alpha(1-6\mathcal{B}^2_\alpha+\mathcal{B}^4_\alpha)\nonumber\\
=\pm2(L/\xi_\omega)(\mathcal{B}_\alpha+\mathcal{B}^3_\alpha),\quad
\alpha=R,L
\end{eqnarray}
where $\mathcal{G}$ and $\mathcal{F}$-functions were defined earlier
in Eqs.~\eqref{G} and \eqref{F}. With this at hand, we find
Josephson current in the form
\begin{equation}\label{I-sns-long}
eI(\phi)R_N=128\pi T\sin\phi \sum_{\omega>0}\frac{L}{\xi_\omega}
\mathcal{B}_{R}(\omega)\mathcal{B}_{L}(\omega)e^{-L/\xi_\omega}
\end{equation}
which is applicable in the broad range of temperatures
$\varepsilon_{Th}\ll T\lesssim|\Delta_\lambda|$. At lowest
temperatures $T\ll\varepsilon_{Th}$, the current-phase relationship
in Eq.~\eqref{I-sns-long} deviates from being simply sinusoidal
because a separable approximation for $F$-function fails to account
properly for the proximity-induced Thouless gap. Unfortunately,
analytical calculation of $I(\phi)$ is not possible in this limit,
however one may easily estimate the magnitude of the critical
current as $eI_cR_N\sim\varepsilon_{Th}$. Furthermore, it is
expected that $I(\phi)$ will be nonanalytical function near
$\phi=\pi$ since the proximity gap closes at that point while the
current is proportional to its derivative
$I(\phi)\propto\partial_\phi\varepsilon_g(\phi)$.

\begin{figure}
  \includegraphics[width=8cm]{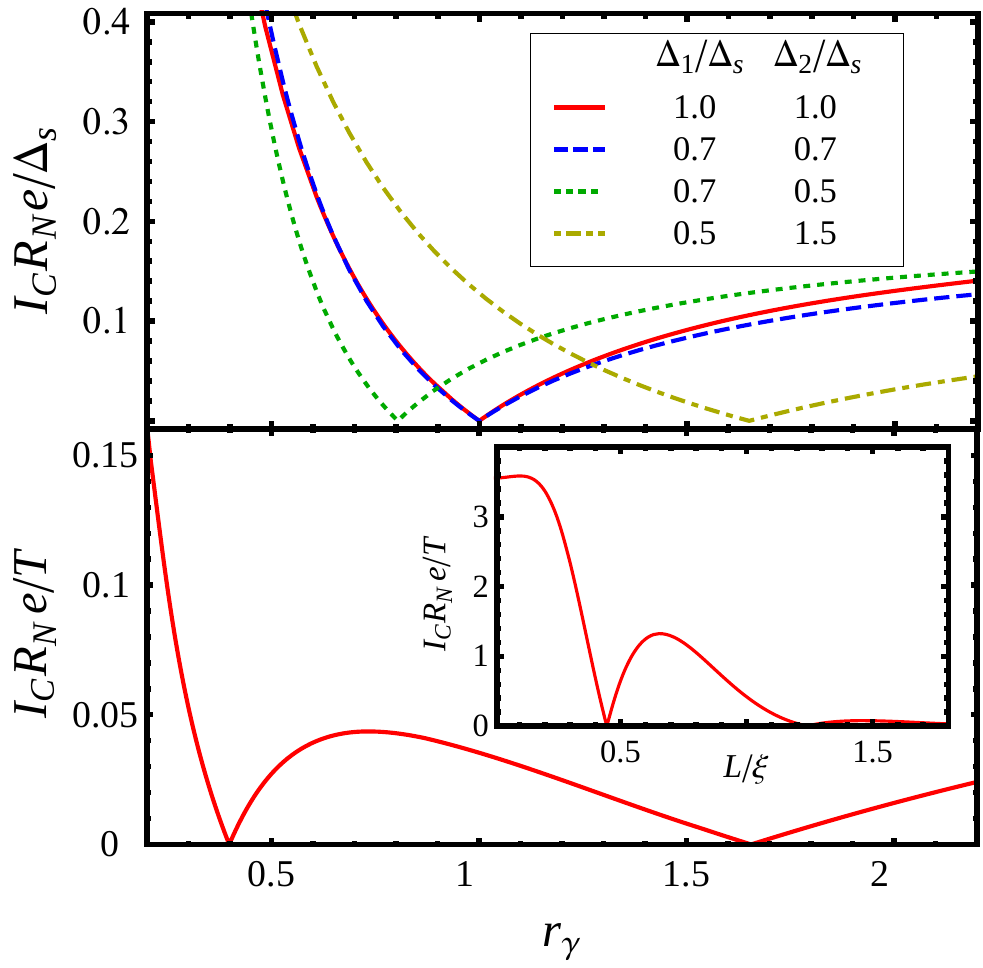}\vspace{-.25cm}\\
  \caption{(Top) Critical current for $s|n|s^{+-}$ junction vs
  boundary resistance mismatch $r_\gamma$. In the inset, $\Delta_s$ labels the gap of
  an ordinary superconductor, while $\Delta_{1,2}$ are the two gaps of an $s^{+-}$
  superconductor. The other parameters are $T=0.5\Delta_s$,
  $\gamma_1=\gamma_s=5$, and $L/\xi_{\pi T}=2$.
  (Bottom) Critical current for $s|f|s^{+-}$ junction vs
  boundary resistance mismatch $r_\gamma$ for $T=0.3\Delta_s$, $\Delta_1=0.5\Delta_s$,
  $\Delta_2=1.5\Delta_s$, and $h=3\Delta_s$.}\label{Fig-I-SFS}
\end{figure}

We plot in Fig.~\ref{Fig-I-SFS} the critical current from
Eq.~\eqref{I-sns-long} as a function of the ratio between the
interface barriers for each band $r_\gamma=\gamma_1/\gamma_2$. It is
well known that for the conventional superconductors, $I_c$ decays
monotonously with $r_\gamma$, however, for $s|n|s^{+-}$ junction
critical current displays clear $0-\pi$
switching.~\cite{Chen-PRL09,Linder-PRB09,Berg-PRL11,Vakaryuk} This
effect is magnified in the presence of ferromagnetic layer.
Including the exchange field $h$ in Eq.~\eqref{Usadel-Eq1} as
$\omega\to\omega+ih\sign(\omega)$, but ignoring spin-flip and
spin-orbital scattering, we find from the linearized Usadel
equations the current in $s|f|s^{+-}$ junction $I(\phi)=I_c\sin\phi$
with
\begin{equation}\label{I-sfs-long}
eI_cR_N=4\pi T\!\!\sum^{\infty}_{\omega=-\infty}\!\!
\frac{(L/\xi_{|\omega|})(\mathcal{F}_R\mathcal{F}_L/\mathcal{G}_R\mathcal{G}_L)/\cosh(L/\xi_{|\omega|})}
{(1+\Gamma^2_\omega)\tanh(L/\xi_{|\omega|})+\Gamma_\omega\mu_\omega}
\end{equation}
where
$\Gamma_\omega=L/2\xi_{|\omega|}\sqrt{\mathcal{G}_R\mathcal{G}_L}$
and
$\mu_\omega=(\mathcal{G}_R+\mathcal{G}_L)/\sqrt{\mathcal{G}_R\mathcal{G}_L}$.
Equation~\eqref{I-sfs-long} is the generalization of the Buzdin
formula~\cite{Buzdin-PRB03} for the multi-band case. Lower panel of
Fig.~\ref{Fig-I-SFS} shows enhanced $0$-$\pi$ oscillations of the
critical current as a function of $r_\gamma$, which displays two
zero points. Such a peculiar feature is due to the combination of a
ferromagnet and $s^{+-}$ superconductor.

\begin{figure}
  \includegraphics[width=8cm]{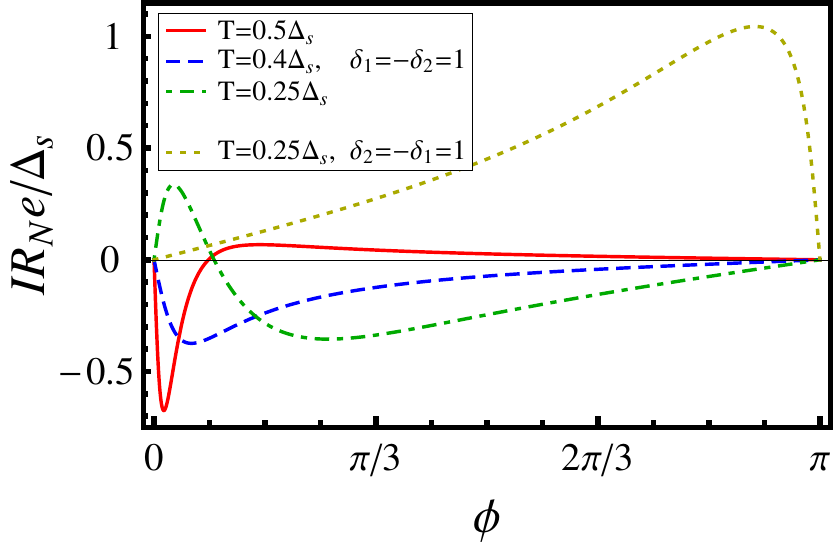}\vspace{-.25cm}\\
  \caption{Josephson current-phase relationship for short $s|n|s^{+-}$ junction for
  gaps $\Delta_1= 0.5\Delta_s$, $\Delta_2=1.5\Delta_s$, and
  interface parameters $\gamma_s=2$, $\gamma_1=2.4$, $\gamma_2=1.8$.}\label{Fig-I-ph}
\end{figure}

In the Josephson junction with extremely low barrier transparency
when $\gamma_\lambda\gg1$, one can circumvent the need of solving
Usadel equation in the wire since current is largely determined by
the interface. Superconductive phase $\phi$ changes discontinuously
at the barriers and stays nearly zero within the interior of the
wire while Green's function phase $\theta$ is approximately
constant. Since $J_\omega\propto\gamma^{-1}\ll1$, then to the
leading order, one can set $\chi_B=0$ in Eq.~\eqref{Boundary-Cond1}
and $\partial_x\theta=0$ in Eq.~\eqref{Boundary-Cond2}. These
approximations allow to find the spectral current,
\begin{equation}
J_\omega=\mathcal{F}^2\sin\phi
\left[\mathcal{G}^2+\mathcal{F}^2\cos^2\frac{\phi}{2}\right]^{-1/2},
\end{equation}
where we assumed $s^{+-}|n|s^{+-}$ geometry. For the equal gaps case
$\Delta_1=\Delta_2\equiv\Delta_s$, above the spectral current
$J_\omega$ leads to the Josephson current-phase relationship from
Eq.~\eqref{I-def}:
\begin{eqnarray}
&&eI(\phi)R_N=\frac{\mu^2(1+r_\gamma)}{2\gamma_1}\Delta_s\sin
\phi\nonumber\\
&&\times\int^{\Delta_s}_{\mu\Delta_s\cos\frac{\phi}{2}}
\frac{\tanh(\varepsilon/2T)d\varepsilon}{\sqrt{\Delta^2_s-\varepsilon^2}
\sqrt{\varepsilon^2-\mu^2\Delta^2_s\cos^2\frac{\phi}{2}}}
\end{eqnarray}
where $r_\gamma=\gamma_1/\gamma_2$ and
$\mu=(1-r_\gamma)/(1+r_\gamma)$. Interestingly, in the
zero-temperature limit, even the remaining energy integral can be
completed in the closed form, such that we find a Josephson current
\begin{equation}
eI(\phi)R_N=\frac{(\gamma_2-\gamma_1)^2\Delta_s\sin\phi}{\gamma_1\gamma_2(\gamma_1+\gamma_2)}
K\!\left[1-\left(\frac{\gamma_2-\gamma_1}{\gamma_2+\gamma_1}\right)^2\!\!
\cos^2\frac{\phi}{2}\right],
\end{equation}
where $K(x)$ is the complete elliptic integral of the first kind. In
the completely symmetric case (with equal transparencies for both
bands), the Josephson current vanishes, since the $\pi$-shifted
bands drive it in the opposite directions. However, including
interband scattering effects may result in additional nonvanishing
contributions.~\cite{Yerin}

Analytical results are also possible for arbitrary transparencies
but for the short junctions when $L\ll\xi_{\omega=\Delta_\lambda}$.
In this case, Usadel equations are dominated by the gradient terms.
Despite the fact that they remain coupled and nonlinear, all
integrations can be completed in the closed form.~\cite{AL-I} By
using the first integral of Eq.~\eqref{Usadel-Eq2},
$J_\omega=\sin^2\theta\partial_x\chi$ and excluding $\partial_x\chi$
from Eq.~\eqref{Usadel-Eq1}, one finds
\begin{equation}
\partial^2_x\theta=\frac{J^2_\omega\sin(2\theta)}{2\sin^4\theta}.
\end{equation}
This nonlinear differential equation is solved by
\begin{equation}
\cos[\theta(x,\omega)]=\cos\theta_0\cos[J_\omega(x-x_0)/\sin\theta_0].
\end{equation}
Knowing $\theta(x,\omega)$, one can now calculate the second
integral of Eq.~\eqref{Usadel-Eq2},
\begin{equation}
\chi-\chi_0=J_\omega\int^{x}_{x_0}\frac{dx}{\sin^2\theta},
\end{equation}
which reads
\begin{equation}
\sin\theta_0\tan[\chi(x,\omega)-\chi_0]=\tan[J_\omega(x-x_0)/\sin\theta_0].
\end{equation}
Having found explicit solutions for the Green's functions, the
boundary problem for the integration coefficients can be reduced to
solving three algebraic equations:
\begin{eqnarray}
J_\omega=\mathcal{F}_L\sin\theta_B\sin(\chi_B+\phi/2),\\
J_\omega=\mathcal{F}_R\sin\theta_B\sin(\phi/2-\chi_B),\\
\mathcal{F}_L\cos\theta_B\cos(\chi_B+\phi/2)-\mathcal{G}_L\sin\theta_B\nonumber\\
=\mathcal{F}_R\cos\theta_B\cos(\chi_B-\phi/2)-\mathcal{G}_R\sin\theta_B.
\end{eqnarray}
These expressions finally lead us to the Josephson current-phase
relationship in the form
\begin{equation}\label{I-sns-short}
eI(\phi)R_N=8\pi
T\!\!\sum_\omega\!\frac{\mathcal{A}(\phi)\sin\phi}{\mathcal{F}^{-1}_R+\mathcal{F}^{-1}_L}\!
\left[\mathcal{A}^2(\phi)+\frac{(\mathcal{G}_R-\mathcal{G}_L)^2}
{(\mathcal{F}_R-\mathcal{F}_L)^2}\right]^{-\frac{1}{2}},
\end{equation}
where
\begin{equation}
\mathcal{A}(\phi)=\left[\cos^2\frac{\phi}{2}+
\frac{(\mathcal{F}_R-\mathcal{F}_L)^2}
{(\mathcal{F}_R+\mathcal{F}_L)^2}\sin^2\frac{\phi}{2}\right]^{-1/2}.
\end{equation}
A representative feature of Eq.~\eqref{I-sns-short} is that
$I(\phi)$ switches its sign in between $\phi=0$ and $\phi=\pi$ as
shown in Fig.~\ref{Fig-I-ph}. This implies that the free energy of
$s|n|s^{+-}$ junction has two minima and such junction may be used
as the phase inverted in superconducting digital circuits. Such
feature, however, is not unique for $s^{+-}$ superconductors and can
be realized in other complex hybrid circuits with ordinary
materials.~\cite{Golubov-RMP04}

In summary, we have studied the density of states and Josephson
current in mesoscopic circuits with unconventional $s^{+-}$
superconductors. We find that tunneling spectra have distinct
fingerprints of the sign changing symmetry of the underlying
superconductive order parameter induced by the proximity effect.
Furthermore, the critical current exhibits a robust $\pi$ junction
even in the absence of the ferromagnetic layer. The Josephson
current-phase relationship itself is not indicative of $s^{+-}$
symmetry due to the sensitivity to parameters defining the junction.

We would like to thank Maxim Vavilov for useful discussions,
Valentin Stanev for correspondence regarding
Ref.~\onlinecite{Koshelev-EPL11}, and Norman Birge for reading and
commenting on the paper. This work was supported by Michigan State
University.

\end{document}